\documentstyle[preprint,eqsecnum,aps,epsfig]{revtex}
\tightenlines

\begin{document}
\draft
\preprint{CAVENDISH-HEP-99-15}
\title{Mass Effects in Bose-Einstein Correlations
}
\author{Mark Smith}
\address{
University of Cambridge, Cavendish Laboratory,\\
Madingley Road, Cambridge, England, CB3 0HE}
\date{\today}
\maketitle
\begin{abstract}
Bose-Einstein symmetrization can lead to correlations between
out going identical particles which reflect
the space-time extent of the collision process.  At LEP and LEPII 
these correlations have been studied as a function of the single variable 
$Q=\sqrt{-(p_{1}-p_{2})^{2}}$.  Assuming a simple form for the correlation
function the experiments find source radii dependent on the hadron mass.
In this note, I point out that such effects can arise from purely kinematic 
considerations, although these are unlikely to explain the observed
effects completely.
\end{abstract}

\narrowtext

\section{Introduction}
\label{sec:level1}

Bose-Einstein correlations are of interest to the LEP community as a source 
of systematic error in W-mass determinations
(for example \cite{kunszt96,workgrp,llonblad95,kart97}),
and also to the Heavy Ion 
community (see \cite{NA49,NA35})
as a valuable measurement tool for the size of the fireball
created in heavy ion collisions (eg. \cite{heinz96} and refs therein).

Due to limited statistics,
measurements of BE enhancement at LEP are often made in terms of a single 
variable $Q = \sqrt{-(p_{1}-p_{2})^{2}}$ where $p_{1}$ and $p_{2}$ are the
4-momenta of a pair of out going bosons.  The enhancement is described
by a correlation function which introduces an extra parameter, the 
HBT (Hanbury-Brown-Twiss \cite{HBT}) radius. 
It has been observed \cite{bialas99} that the radii measured 
depend on the mass of the boson being used.  In general the lighter 
bosons tend to give larger radii.   
In this section I describe the construction of the correlation function
as a function of $Q$ and its relation to the single particle Wigner
distribution.
In section 2  I use a simple Gaussian model for the Wigner density, 
and discuss the correlation function in various limits.
In section 3 numerical results are presented for particles of 
various different masses and sources of different sizes.  
I conclude with a discussion of how far these results can go to 
explain the observed 
mass dependence of experimental Bose-Einstein parameters.

The {\it correlation function\/} can be defined as
\begin{equation}
C(Q) = \frac{\rho_{2}(Q)}{\rho_{1}(Q)}\label{corrq},
\end{equation}
where $\rho_{2}(Q)$ is the density of particle pairs with invariant
momentum
separation $Q$, and $\rho_{1}(Q)$ is the same density measured with 
respect to a reference distribution.   
The densities $\rho_{1}$ and $\rho_{2}$ can be
defined in terms of the 
single- and two-particle momentum densities ($P_{1}$ and $P_{2}$
respectively)  according to
\begin{equation}
\rho_{1}(Q)  =  \int d^{3}{\bf p}_{1} \; d^{3}{\bf p}_{2} \;
P_{1}({\bf p}_{1}) P_{1}({\bf p}_{2})
\delta (\tilde{Q}({\bf p}_{1},{\bf p}_{2})-Q) \label{rho1}
\end{equation}
\begin{equation}
\rho_{2}(Q) = \int d^{3}{\bf p}_{1} \; d^{3}{\bf p}_{2} \;
P_{2}({\bf p}_{1},{\bf p}_{2})\delta 
(\tilde{Q}({\bf p}_{1},{\bf p}_{2})-Q)\label{rho2},
\end{equation} 
where the function $\tilde{Q}({\bf p}_{1},{\bf p}_{2})$ is just
$\sqrt{-(p_{1}-p_{2})^{2}}$. 

The single-particle momentum distribution is given in terms
of the one-particle Wigner distribution $S(x,p)$,
\begin{equation}
P_{1}({\bf p}_{1}) = \int d^{4}x \; S(x,p)|_{p^{2}=m^{2}}.
\end{equation}
It has long been known that 
under rather general assumptions the two-particle distribution can 
also be
related to the one-particle Wigner density \cite{shuryak73}
\begin{equation}
P_{2}({\bf p}_{1},{\bf p}_{2})=P_{1}({\bf p}_{1}) P_{1}({\bf p}_{2})
+\left| \int d^{4}x \; S(x,K) e^{-i q \cdot x} \right|^2,
\label{twodist}
\end{equation}
where $K = \frac{1}{2} (p_{1}+p_{2})$ and $q=p_{1}-p_{2}$.  
It is thus possible, given the source 
distribution, to compute the densities $P_{1}$ and $P_{2}$ and
hence $\rho_{1}$ and $\rho_{2}$, then to 
construct explicitly the correlation function $C(Q)$.  The correlation
function $C(Q)$ may depend on the particle masses through the integration
limits in equations (1.2) and (1.3), even when there is no explicit
mass dependence in the source function $S$.  In the next section
I examine this implicit dependence for a Gaussian source.

\section{Simple Model}

To illustrate the kinematical mass effects it is best to take a simple
model.  Consider an instantaneous, uncorrelated source, Gaussian in
both space and momentum.
The  normalised source distribution is given by
\begin{eqnarray}
S(x,p) & = & \frac{1}{(2 \pi R_{0}^{2})^{\frac{3}{2}}}
\exp\left(-\frac{|{\bf x}|^{2}}{2R_{0}^{2}}\right)\times
\nonumber 
\\ 
 & & \frac{1}{(2 \pi P_{0}^{2})^{\frac{3}{2}}}
\exp\left(-\frac{|{\bf p}|^{2}}{2P_{0}^{2}}\right) 
\delta(x^{0}).\label{source}
\end{eqnarray}
The parameters $R_{0}$ and
$P_{0}$ are the typical size of the source in position and 
momentum space, respectively. The single and two particle distributions
can now be obtained using the formulae of the previous section:
\begin{eqnarray}
P_{1}({\bf p}) & = & \frac{1}{(2 \pi P_{0}^{2})^{\frac{3}{2}}}
\exp \left( -\frac{|{\bf p}|^{2}}{2P_{0}^{2}} \right)\label{single} 
\\
P_{2}({\bf p}_{1},{\bf p}_{2}) & = & P_{1}({\bf p}_{1})
P_{1}({\bf p}_{2}) + \nonumber \\
& & \frac{1}{(2 \pi P_{0}^{2})^{3}}
\exp\left( -\frac{|{\bf K}|^{2}}{4P_{0}^{2}}-
R_{0}^{2} | {\bf q}|^{2} \right)\label{double},
\end{eqnarray}
where ${\bf K}=\frac{1}{2}({\bf p}_{1}+{\bf p}_{2})$ and
${\bf q}=({\bf p}_{1}-{\bf p}_{2})$.
Note that in this simple example the fully differential particle
distributions do {\bf not} depend on the particle masses.

Inserting Eqs.\ (\ref{single},\ref{double}) into 
Eqs.\ (\ref{rho1},\ref{rho2}) and then constructing
Eq.\ (\ref{corrq})
gives the expression for the correlation function
\widetext
\begin{equation}
C(Q) = 1 + \frac{
\int d^{3} {\bf p}_{1} \; d^{3} {\bf p}_{2} \;
\exp\left( -\frac{|{\bf p}_{1}+{\bf p}_{2}|^{2}}{4P_{0}^{2}}-
R_{0}^{2} | {\bf p}_{1} - {\bf p}_{2}|^{2} \right)
\delta (\tilde{Q}({\bf p}_{1},{\bf p}_{2})-Q)}
{\int d^{3} {\bf p}_{1} \; d^{3} {\bf p}_{2} \;
\exp \left( -\frac{|{\bf p}_{1}|^{2}}{2P_{0}^{2}} \right)
\exp \left( -\frac{|{\bf p}_{2}|^{2}}{2P_{0}^{2}} \right)
\delta (\tilde{Q}({\bf p}_{1},{\bf p}_{2})-Q)}.\label{correqn}
\end{equation}
\narrowtext
Due to the spherical symmetry of the source most of the angular integrations
are trivial.  The $\delta$-function constraint can be satisfied with the
angle between ${\bf p}_{1}$ and ${\bf p}_{2}$; define the cosine of this
angle to be $x$.  The integrations may then be re-written,
\begin{equation}
d^{3} {\bf p}_{1} \; d^{3} {\bf p}_{2} 
\delta (\tilde{Q}({\bf p}_{1},{\bf p}_{2})-Q)
\longrightarrow
p_{1}\;dp_{1}\, p_{2}\;dp_{2} \Theta 
(1-x) \Theta(1+x)\label{int}
\end{equation}
where $ p_{i} = |{\bf p}_{i}|, \;\;\;\; i=1,2$.  Constants independent
of $p_{1}$ and $p_{2}$ have been dropped as these cancel in the 
ratio (\ref{correqn}). The variable $x$ as a function of $p_{1}$ and
$p_{2}$ is given by
\begin{equation}
x=\left( \frac{
\sqrt{(p_{1}^{2} + m^{2})(p_{2}^{2} + m^{2})} - m^{2}}
{p_{1}p_{2}} \right) - \frac{Q^{2}}{2 p_{1}p_{2}},
\end{equation}
where $\Theta(z)$ is the Heaviside (theta) function.  Notice that mass
dependence has now crept into $C(Q)$ via the integration limits.

The theta-functions in Eq.\ (\ref{int}) give rise to a complicated
integration region in the $p_{1}-p_{2}$ plane which depends on the
mass of the bosons.  The integration region can be summarized as
\begin{equation}
p_{2}  >  
\left|
p_{1}\left( 1+\frac{Q^{2}}{2m^{2}} \right) - Q
\sqrt{\left( 1 + \frac{p_{1}^{2}}{m^{2}}\right)
\left(1+\frac{Q^{2}}{4m^{2}}\right)} \right|, 
\end{equation}
\begin{equation}
p_{2}  < 
p_{1}\left( 1+\frac{Q^{2}}{2m^{2}} \right) + Q 
\sqrt{\left( 1 + \frac{p_{1}^{2}}{m^{2}}\right)
\left(1+\frac{Q^{2}}{4m^{2}}\right)}.
\end{equation}
This takes on simple forms in the following limits
\begin{eqnarray}
\frac{Q}{m} \longrightarrow \infty & \,:\,\,\,\,\,\,\, 
& p_{2} > \frac{Q^{2}}{4 p_{1}}
\,\,\,\,\,\,\,\,\,\,\,\, \mbox{   (the massless limit)} \\
\frac{Q}{m} \longrightarrow 0 & \,:\,\,\,\,\,\,\,
& p_{1} \left(1-\frac{Q}{m}\right) <
p_{2} < p_{1} \left(1+\frac{Q}{m}\right) \nonumber \\
 & & \;\;\;\;\;\mbox{  where  } 
p_{1},p_{2} \gg m.\label{limit2} \\
\frac{Q}{m} \longrightarrow 0 & \,:\,\,\,\,\,\,\,
& |p_{1}-Q| < p_{2} < p_{1}+Q\;\;\;\;\;\; \nonumber \\
 & & \;\;\;\;\;\;\mbox{where  }
p_{1},p_{2} \ll m
\end{eqnarray}

Now the BE enhanced region of the $p_{1}-p_{2}$ plane lies
on a band of thickness $\sim R_{0}^{-1}$
centered around 
$p_{1} \approx p_{2}$, while the reference density lies largely in the 
region $p_{1},p_{2} \leq {\mathcal O} (P_{0})$.  
Figures 1 and 2 show the enhanced region and the integration bounds
for $Q < m$ and $Q > m$.
It is then clear that
for any $Q \gg m$ the enhanced region occupies a fraction 
${\mathcal O}([R_{0}P_{0}]^{-1})$ of the populated region.  
This factor must always 
be less than one by the uncertainty principle and so the strength of the 
correlation is reduced.  In fact it is possible to show that in the massless
limit, the intercept of the correlation function at $Q=0$ occurs at
\begin{eqnarray}
C(Q \approx 0) &  = & 1+
\left( \frac{ \xi - (1-\xi^{2})\tan^{-1}(\xi)}
{\xi^{3}} \right) \nonumber \\
 & \approx & 1+\frac{\pi}{2\xi}\;\;\;\;\; \mbox{as } \xi \longrightarrow
\infty,\label{corrsuppress}
\end{eqnarray}
where $\xi=2R_{0}P_{0}$.

The opposite limit $Q \ll m$ is qualitatively different.  One can
see from Eq.\ (\ref{limit2}) that as $Q \longrightarrow 0$ with
$Q \ll m$, the integration region is squeezed into a narrow wedge
along the diagonal of the $p_{1}-p_{2}$ plane (the integration region
in figure 1 is approaching this limit).  
If the inequality
\begin{equation}
\frac{P_{0} Q}{m} < \frac{1}{R_{0}}\;\;\;\;\;\;\;
\Longleftrightarrow \;\;\;\;\;\;\;
Q < \frac{m}{R_{0}P_{0}}
\end{equation}
is satisfied then the integration region only contains the enhanced 
region and the correlation function can approach the expected value 
of $C(Q \approx 0)=2$.  In fact in the limit that the mass is large 
compared to all other scales, the correlation function approaches
the `naive' form
\begin{equation}
C(Q)=1+\exp(-Q^2 R^{2}_{class.}) \label{non-rel}
\end{equation}
where
\begin{equation}
R^{2}_{class.}=R_{0}^{2}\left(1-\frac{1}{(2 R_{0}P_{0})^{2}}
\right).
\end{equation}

\section{Numerical Results}

In the case of an arbitrary mass the integrations must be done
numerically.  The correlation function shows different behaviour
depending on the values of the mass and source parameters. 
The results of numerical evaluation of eqn.(\ref{correqn})  are 
illustrated in figures (3,4).
The typical momentum is set to the value
\begin{equation}
P_{0} = 1\;\; \mbox{GeV},
\end{equation}
in order to be comparable with the observed pion spectrum at the
Z peak.  This implies $R_{0} > 0.1$ fm which is consistent with 
all current estimates of the source size from BE Correlations.

Figure 3 shows the computed correlation function for a source 
of size $0.2$ fm, and for three values of the mass roughly
corresponding to the pion, kaon and $\Lambda$ baryon\footnote{
Here I plot positive correlations for the antisymmetric 
$\Lambda \Lambda$ spin state for
ease of comparision.  The $\Lambda$ is, of course, a fermion and
after spin averaging displays anticorrelations,
which lead to a suppression
at small relative $Q$.  This does not affect discussion of the
HBT radius} 
masses.  This corresponds
to near the limit of the uncertainty principle ($R_{0}P_{0}=\frac{1}{2}$).
As implied in 
eq(\ref{corrsuppress}) the correlation is more strongly suppressed
in the region $Q > m$ if $R_{0}P_{0}$ is large.  This is, in fact, quite
general and the opposite is true; when $R_{0}P_{0}$ is close to the 
uncertainty limit the correlation function is only weakly suppressed.
This is because when $R_{0}^{-1} \sim P_{0}$, the enhanced region
extends over {\em all} the populated phase space (see discussion
and figures from previous section) and averaging over the phase space
causes no dilution.

The opposite limit ($R_{0}P_{0}$ large) is explored in figure 4.  Here 
the source radius is $R_{0}=0.8$ fm which takes $R_{0}P_{0}$
sufficiently far from the Heisenberg limit to give strong suppression.
The effect is particularly noticeable for the pion correlation which
becomes suppressed by a roughly a factor of $4\sim R_{0}P_{0}$ in the 
region $Q > m$.  This leads to typically non-Gaussian shapes for the
pion (and other low mass) correlation functions.  The suppression
differentiates between the kaon and $\Lambda$ correlations leading
to a hierarchy of effective radii, $R_{\pi} > R_{K} > R_{\Lambda}$.

From figures (3,4) one can see that smaller masses give characteristically 
sharper correlation functions as compared to heavier bosons.  One can
define an effective radius for the source (somewhat arbitrarily)
in terms of the width at $e^{-1}$ of the maximum
of the correlation function.  This results in the 
estimates shown in table 1 ( see \cite{pions} and 
\cite{kaons} for pion and kaon measurements respectively).

The reader may wonder if perhaps some intermediate source radius
would provide a fit to the experimental radii.  This is not the case.  If 
one insists on fitting the source size measured by $\Lambda$ correlations
one is forced to a source of no larger than  $\sim 0.2$ fm.  
The constraint that $P_{0} \sim 1.0$ GeV now 
leaves $R_{0} P_{0}$ too small to provide the 
suppression necessary to fit the kaon and pion HBT radii.  On the other hand,
a source with $R_{0}P_{0} > 1$ leads to natural radii for pion and
kaon correlations of order $R_{\pi} \sim m_{\pi}^{-1}$ and $R_{K}
\sim m_{K}^{-1}$ which are larger than experimentally observed.  An 
intermediate choice for $R_{0}$ leads to distorted correlation
functions which are partially suppressed for $Q>m$.  These are difficult
to parameterise in terms of a single radius. 

It is interesting to note that the correlation function for $R_{0}P_{0} \gg 1$
and $m < P_{0}$ can be rather well approximated by the expression:
\begin{equation}
C(Q) \approx 1+ \exp (-R_{class}^{2} Q^{2}) \left(
\frac{y+(1+y^{2})\tan^{-1}(y)}{2y(1+y^{2})} \right)
\left(\frac{1+\frac{3Q}{4m}+\frac{Q^{2}}{m^{2}}}
{1+\frac{Q}{m}}
\right),
\end{equation}
where
\begin{equation}
y=\frac{Q R_{0}P_{0}}{m}.
\end{equation}
The suppression term multiplying the Gaussian tends to unity as
$Q\rightarrow 0$, and approaches the value $\frac{\pi m}{2\xi Q}$
as $Q \rightarrow \infty$,
where $\xi = 2 R_{0} P_{0}$ as given in
eqn.(\ref{corrsuppress}).

\section{Conclusions}

I have shown that the averaging over phase-space to produce a correlation
function of a single variable (Q) induces non-trivial, mass dependent
distortions on the correlation function.
It has been demonstrated that these distortions are such that correlations
between smaller mass bosons are inherently shorter range in Q and
lead to larger apparent radii.  The effect is greatest away
from the Heisenberg limit.

For the physically reasonable value of $P_{0}=1.0$ GeV,
it is not possible to find a source radius $R_{0}$ such that the 
kinematic averaging simultaneously fits the experimentally
measured HBT radii in pion, kaon and $\Lambda$ baryon correlations.
As can be seen from table 1, away from $R_{0}P_{0} \sim 1$, averaging
over phase-space can induce a mass dependence of the HBT radii of 
the same order of magnitude as experimentally observed
(but not with the same absolute values).

Other effects are therefore necessary to bring theory into line with 
experimental data, but the effects of phase-space averaging should
not be neglected.

\acknowledgements

I am grateful to Ulrich Heinz for many valuable discussions.  The
hospitality of the CERN Theory Division during this work is 
gratefully acknowledged.  Research supported in part by the
EU Programme `Training and Mobility of Researchers', Network 
`Quantum Chromodynamics and the Deep Structure of Elementary Particles',
contract FMRX-CT98-0194 (DG 12 - MIHT).

\begin{table}
\caption{Comparison of the radius parameters extracted from the simple
Gaussian model (for $P_{0}=1$ GeV, $R_{0}=0.2$ fm and $R_{0}=0.8$ fm) 
and experimentally measured values.}
\begin{tabular}{|l|c|c|c|} 
Mass / GeV & 0.14 & 0.5 & 1.1 \\ \tableline
(Source with $R_{0}=0.2$ fm) Radius / fm  & 0.32 & 0.29 & 0.22 \\ \tableline
(source with $R_{0}=0.8$ fm) Radius / fm  & 2.8 & 1.2 & 0.9 \\ \tableline
Experimental Radius / fm & 0.9 & 0.6 & 0.2 \\
\end{tabular}
\end{table}

\begin{centering}

\begin{figure}
\epsfig{file=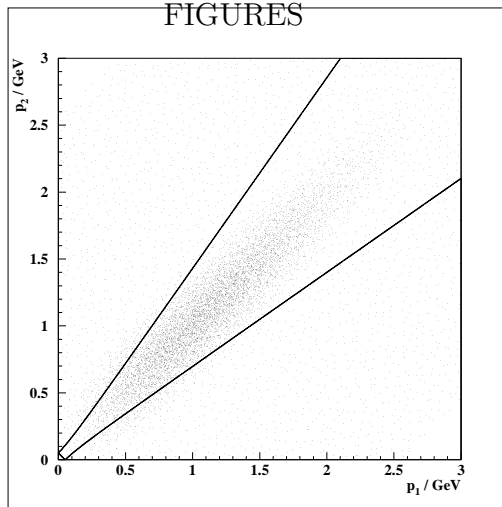,height=6cm, width=6cm,
bbllx=20pt,bblly=143pt, bburx=520pt, bbury=644pt}
\caption{A plot showing the allowed integration region given
the parameters $m=0.14$ GeV, 
$Q=0.05$ GeV,
$R_{0}=0.8$ fm and $P_{0}=1.0$ GeV.  
The scatter shows schematically the enhanced part
of the phasespace.  For $Q < m$ the integration is mainly over the 
enhanced region}
\end{figure}

\begin{figure}
\epsfig{file=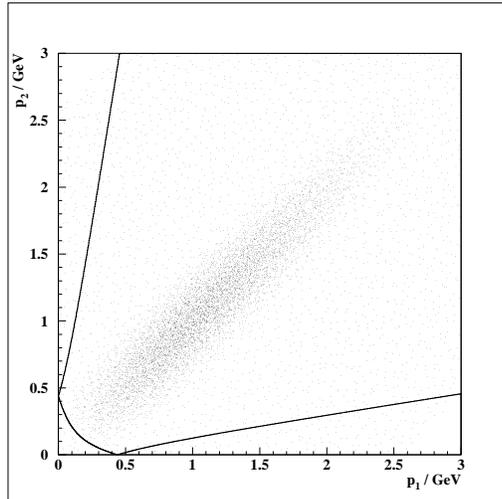,height=6cm,width=6cm,
bbllx=20pt,bblly=143pt,bburx=520pt, bbury=644pt}
\caption{A similar plot to fig 1, but with paramter $Q=0.3$ GeV.
With $Q > m$ the allowed integration region is not restricted to the 
enhanced region and the correlation is diluted.}
\end{figure}

\pagebreak

\begin{figure}
\epsfig{file=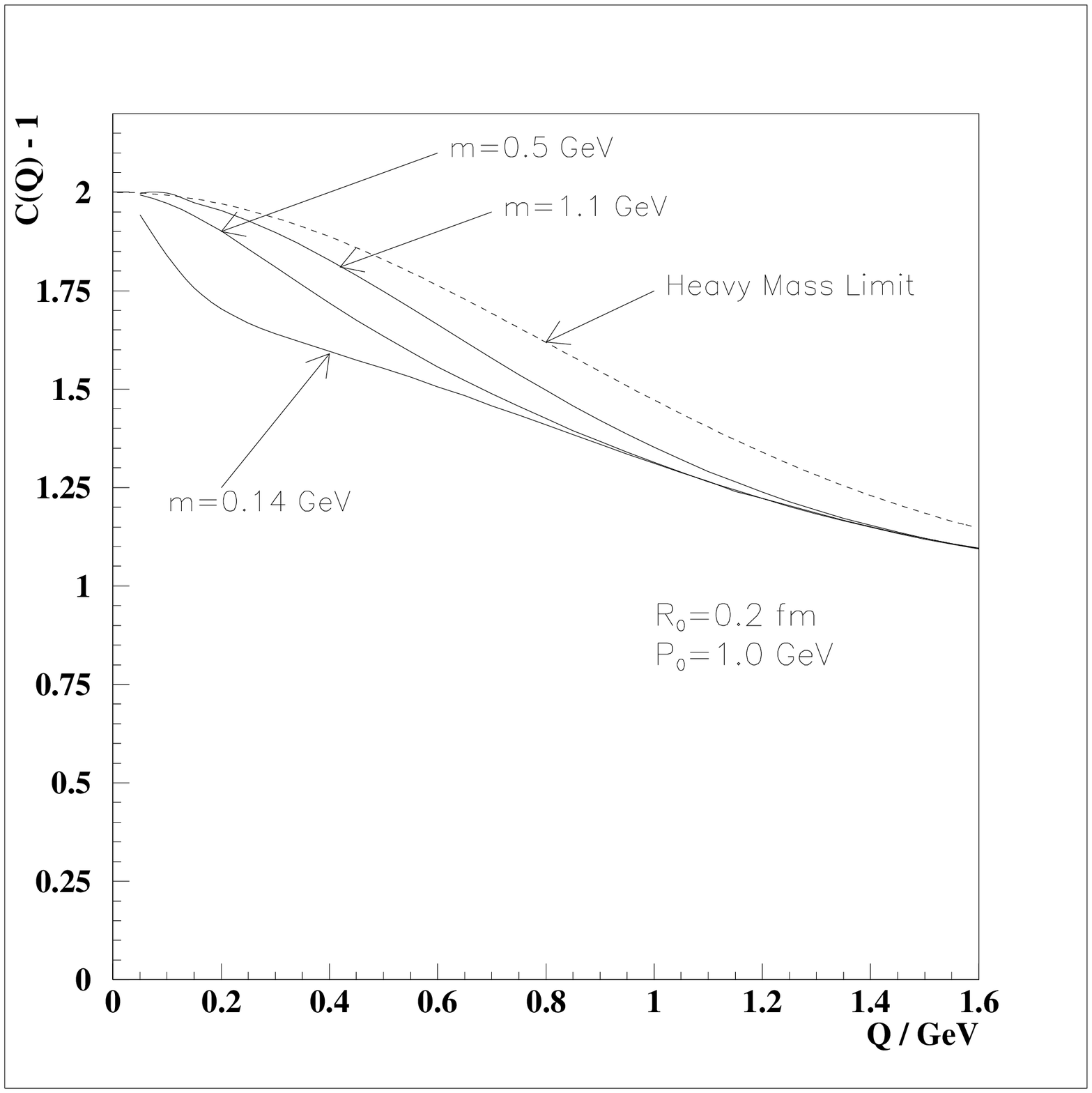,height=6cm, width=6cm,
bbllx=20pt,bblly=143pt,bburx=520pt,bbury=644pt}
\caption{Correlation functions computed for boson masses 
$m_{\pi}\sim0.14$ GeV,
$m_{K}\sim0.5$ GeV 
and $m_{\Lambda}\sim1.1$ GeV.  
The dashed curve shows the correlation function
in the extreme heavy mass limit which corresponds to the naive expectation.
The correlation functions are computed from a Gaussian source of
radius $R_{0}=0.2$ fm and typical momentum $P_{0}=1.0$ GeV,
 $Q=\sqrt{-(p_{1}-p_{2})^2}$ - the invariant momentum separation.}
\end{figure}

\begin{figure}
\epsfig{file=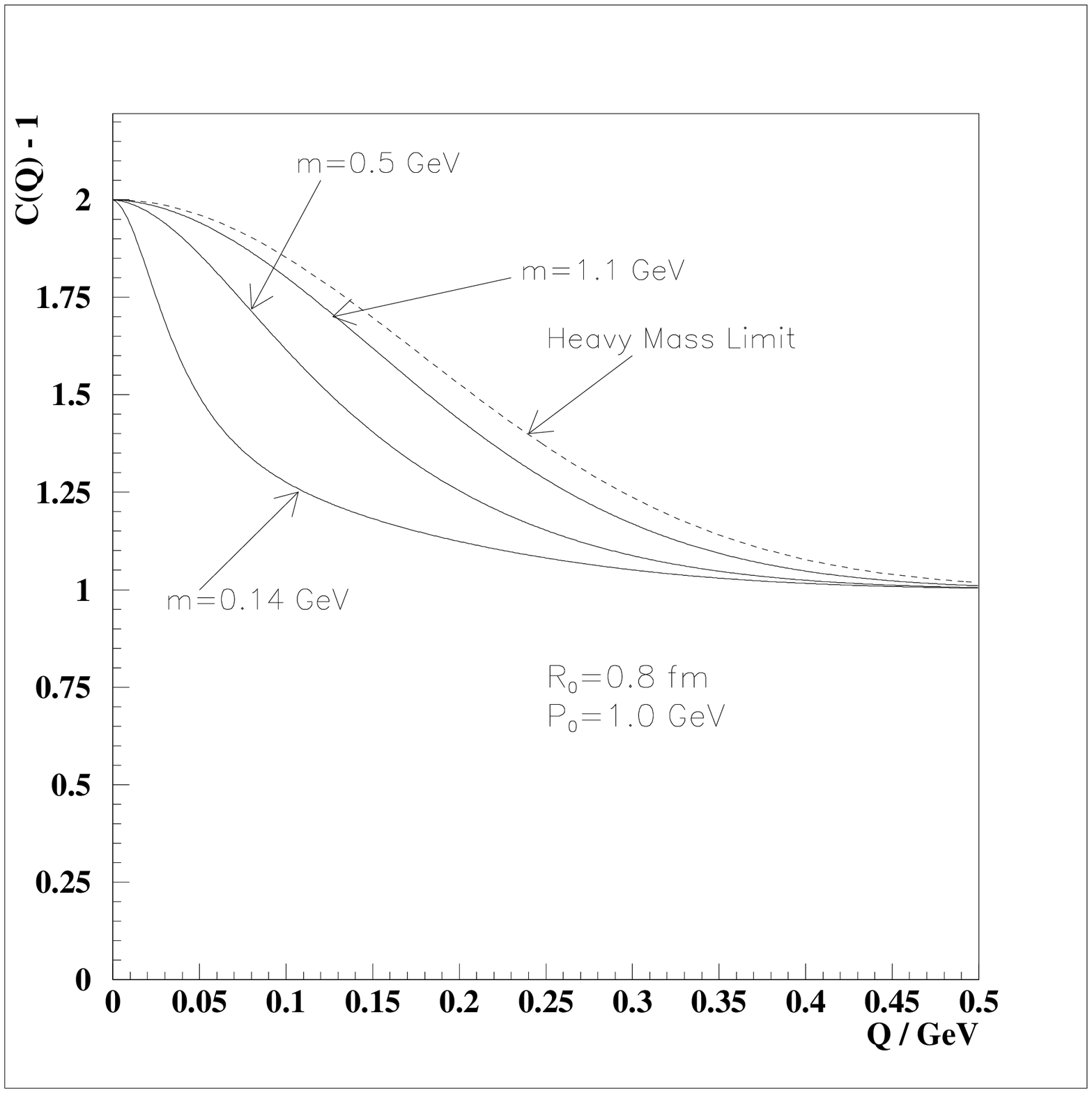,height=6cm,width=6cm,
bbllx=20pt,bblly=143pt,bburx=520pt,bbury=644pt}
\caption{Correlation functions computed for boson masses
$m_{\pi} \sim 0.14$ GeV,
$m_{K}\sim 0.5$ GeV
and $m_{\Lambda} \sim 1.1$ GeV.
The dashed curve shows the correlation function
in the extreme heavy mass limit which corresponds to the naive expectation.
The correlation functions are computed from a Gaussian source of
radius $R_{0}=0.8$ fm and typical momentum $P_{0}=1.0$ GeV,
$Q=\sqrt{-(p_{1}-p_{2})^2}$ - the invariant momentum separation.}
\end{figure}

\end{centering}
\end{document}